\documentclass{article}

\usepackage{arxiv}
\usepackage[utf8]{inputenc} 
\usepackage[T1]{fontenc}    
\usepackage{amsthm,amsmath,amsfonts,amssymb}
\usepackage{hyperref}       
\usepackage{url}            
\usepackage{booktabs}       
\usepackage{nicefrac}       
\usepackage{microtype}      
\usepackage{lipsum}		
\usepackage{graphicx}
\usepackage{natbib}
\usepackage{mathabx}
\usepackage{doi}

\title{Dynamical manifold dimensionality as characterization measure of chimera states in bursting neuronal networks}

\author{ 
\hspace{1mm}Olesia Dogonasheva
\thanks{odogonasheva@gmail.com} \\
	École Normale Supérieure PSL*\\
	Paris, France \\
	HSE University\\
	Moscow, Russia \\
\And
\hspace{1mm}Daniil Radushev \\
        HSE University\\
	Moscow, Russia \\
\And
\hspace{1mm}Boris Gutkin \\
        École Normale Supérieure PSL*\\
	Paris, France \\
\And
\hspace{1mm}Denis Zakharov \\
        HSE University\\
	Moscow, Russia \\
}

\renewcommand{\shorttitle}{Dynamical manifold dimensionality as measure of chimera states}

\hypersetup{
pdftitle={Dimensionality of chimera states},
pdfauthor={Olesia Dogonasheva, Boris Gutkin, Denis Zakharov},
pdfkeywords={Synchronization, Correlation dimension, Spiking neural networks},
}

\begin{document}
\maketitle

\begin{abstract}
Methods that distinguish dynamical regimes in networks of active elements make it possible to design the dynamics of models of realistic networks. A particularly salient example is partial synchronization, which may play a pivotal role in elucidating the dynamics of biological neural networks. Such emergent partial synchronization in structurally homogeneous networks is commonly denoted as chimera states. While several methods for detecting chimeras in networks of spiking neurons have been proposed, these are less effective when applied to networks of bursting neurons. Here we introduce the correlation dimension as a novel approach to identifying dynamic network states. To assess the viability of this new method, we study a network of intrinsically Hindmarsh-Rose neurons with non-local connections. In comparison to other measures of chimera states, the correlation dimension effectively characterizes chimeras in burst neurons, whether the incoherence arises in spikes or bursts. The generality of dimensionality measures inherent in the correlation dimension renders this approach applicable to any dynamic system, facilitating the comparison of simulated and experimental data. We anticipate that this methodology will enable the tuning and simulation of when modelling intricate network processes, contributing to a deeper understanding of neural dynamics.
\end{abstract}

\keywords{Synchronization \and Correlation dimension \and Bursting neurons \and Spiking neural networks}

\section{Introduction}

Neuronal populations process information through flow of action potentials and impulses via electrical and chemical synapses. The resulting interactions in networks give rise to coordinated collective dynamics, exhibiting diverse behaviors such as synchronization, traveling waves, partial clustering, and incoherence \citep{rabinovich2006dynamical}. Partial synchronization, a phenomenon where neurons within a specific cluster display synchronous activity while those in different clusters do not \citep{krupa2014adaptation}, is commonly referred to as chimeras \citep{abrams2004chimera,kuramoto2002coexistence}. These chimera states, characterized by the coexistence of synchronous and non-synchronous clusters in homogeneous networks, are observed across various dynamical systems, including coupled Van der Pol oscillators \citep{bastidas2015quantum, omelchenko2015nonlinearity}, Leaky integrate-and-fire neural networks \citep{olmi2011collective, tsigkri2016multi, tsigkri2017chimeras}, Hindmarsh-Rose neural network \citep{hizanidis2014chimera, hizanidis2016chimera, majhi2017chimera}, network of type-I Morris-Lecar neurons \citep{calim2018chimera}, Belousov–Zhabotinsky reaction \citep{tinsley2012chimera}.

Numerous methods have been proposed to identify chimera states, such as the Strength of Incoherence SI \citep{gopal2014observation} and the ACM measure \citep{dogonasheva2021robust} based on the $\chi^2$ parameter \citep{golomb2001mechanisms}. However, these methods face challenges in adapting to bursting neurons. In this study, we introduce a novel perspective by exploring dimensionality of a dynamical manifold in the phase space.

Network synchronization can be viewed as a dimensional reduction of system dynamics.
In the case of complete (global) synchronization neuronal activity, there exists a symmetrical manifold $\vec{x_1}(t)=\vec{x_2}(t)=...=\vec{x_n}(t)=\vec{x}(t)$, 
while asynchronous elements increase the dimension of the system dynamical manifold. Chimera states, being a form of partial synchronization, can be identified by analyzing the dimension of the dynamical manifold in the phase space.

The fractal \citep{mori1980fractal, peitgen2006functional, russell1980dimension}, information \citep{renyi1959dimension}, and correlation dimensions \citep{grassberger1983characterization} are commonly employed to estimate attractor dimensions in dynamical systems. The fractal dimension, which captures self-similarity at different spatial scales is commonly applied in chaotic volumetric systems. The information dimension serves as a lower bound for the fractal dimension, measuring the system entropy. Finally, the correlation dimension, based on point correlations, assesses the complexity of the dynamical system across spatial scales. In this article, we demonstrate the practical application of the correlation dimension in investigating dynamical systems, employing a network of Hindmarsh-Rose neurons to explore parameter regions associated with the emergence of chimera states.

\section{Methods}

The correlation dimension is rooted in the analysis of dependencies between data points, offering an estimate of the attractor embedding dimension that reflects the dynamical system complexity. The calculation of the correlation dimension employs the Grassberger-Prokaccia algorithm. Consider a time series generated by the system dynamics: $y_1 = y(t_1)$, $y_2 = y(t_2)$, ..., $y_n = y(t_n)$, where $y_i$ represents the vector of dynamical states for each variable of neurons, represented as $N$-dimensional systems, $t$ denotes time. For each pair of time moments, the $L^2$-metric is calculated as:

\begin{equation}
    \rho_{ij} = ||y(t_j) - y(t_i)|| = \sqrt{\sum_{k=1}^{N}\left(y(t_j)_k - y(t_i)_k\right)^2}.
\end{equation}

The correlation integral $C(l)$ is then computed as number of pairs of points separated from each other by a distance of no more than $l$:

\begin{equation}
    C(l) = \lim_{n\to \infty} \frac{1}{n^2}\sum_{i=1}^{n}\sum_{j=1}^{n}\Theta \bigl[ l - \rho_{i, j}\bigr],
\end{equation}

where $\Theta$ is the Heaviside step function. The correlation integral signifies the normalized number of point pairs, with distances less than $l$ contributing to $C(l)$. The correlation dimension, denoted as $d_c$, is then determined by the limit:

\begin{equation}
    d_c = \lim_{l\to 0} \frac{\ln C(l)}{\ln l}, 
    \label{eq:dc}
\end{equation}

\begin{figure}[h]
	\centering
        \includegraphics[width=0.8\textwidth]{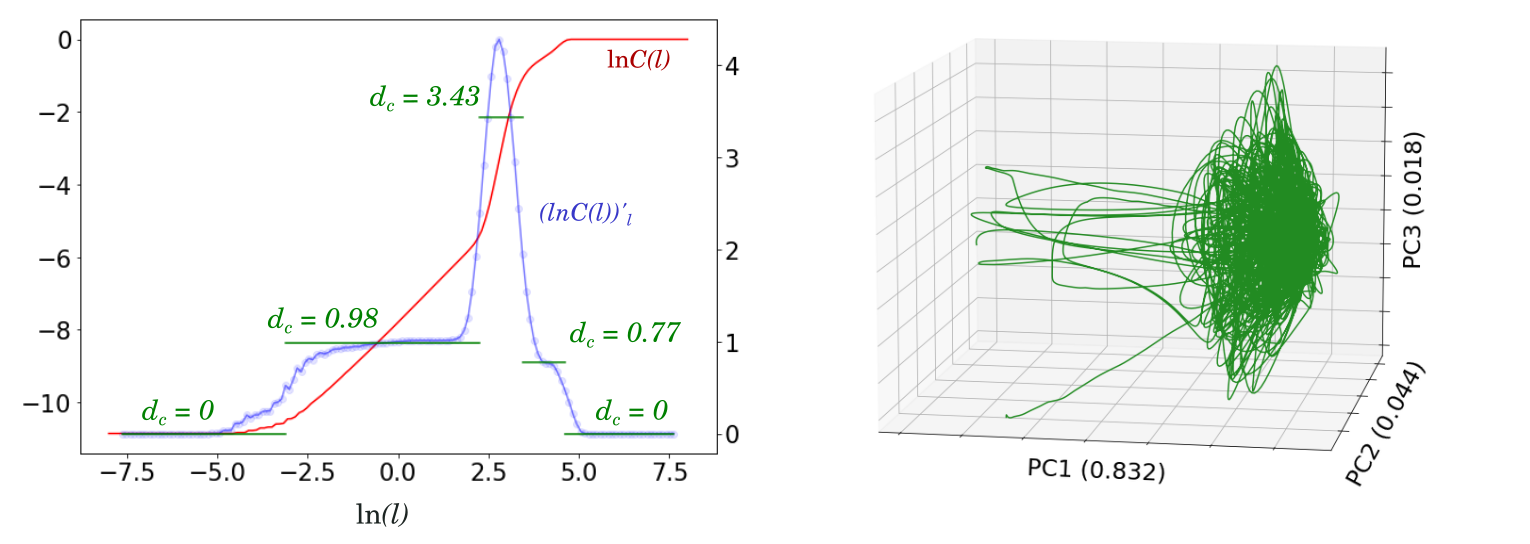} 
	\caption{Left: an example of the existence of different dimensions across spatial scales. The red line is for $\ln(C(l))$, while the blue line corresponds to $\frac{\partial \ln{C(l)}}{\partial \ln{l}}$. The green lines indicate the value of $d_c$ for various scales. These dimensions are calculated as $\frac{\partial \ln{C(l)}}{\partial \ln{l}}$ for specific $ln(l)$, where the derivative exhibits a plateau. 
    Right: Principal Component Analysis (PCA) was performed on the dynamical regime associated with the curves depicted in the left plot.}
	\label{fig:platos}
\end{figure}

Geometrically, $d_c$ corresponds to the tangent of the slope of $\ln C(l)$ with respect to $\ln l$. In the context of simulated dynamics, a manifold is not infinitely densely covered with points but only a certain sample. Consequently, it is impossible to make a limit $l \rightarrow 0$. Therefore, a value of $l$ has to be chosen that is small enough to estimate the limit but large enough to estimate $C(l)$ for this $l$. Analytically, the value of $l$ is found by examining the derivative of the logarithm of the correlation integral, as depicted in Fig.~\ref{fig:platos}. The plateau of the derivative identifies the value of $l$ that can be used for $C(l)$ estimation.

The final step involves establishing thresholds for $d_c$ values to differentiate dynamical regimes in the system. For a dot-like representation of the dynamical manifold in phase space with $d_c = 0$, it implies an absence of oscillations for the considered parameter configuration. Global synchronization corresponds to $d_c = 1$. The boundary between a chimera state and an incoherent regime is typically indistinct. However, in line with the central limit theorem, as the system size $N$ approaches infinity, the behavior of $d_{c_{\infty}} + \frac{k}{\sqrt{N}} + \mathcal{O}(\frac{1}{N})$ is expected, where $k > 0$ is a constant \citep{golomb2001mechanisms}. Thus, as a default boundary for an asynchronous regime, we propose using $d_c = \sqrt{N}$ and suggest adjusting this value based on the specific problem. Threshold information for $d_c$ is summarized in Table~\ref{tab:thr}. Given the imprecision and noise inherent in large complex systems, we opt for introducing bounds on the correlation dimension parameter rather than relying on fixed values. This approach follows the example set by other parameters in defining dynamic regimes.

\begin{table}[h!]
	\caption{Thresholds of $d_c$ for regime separation}
	\centering
	\begin{tabular}{l|r}
		Regime & $d_c$ \\
            \hline
            Absence of oscillations & 0 \\
            Synchronization & (0, 1] \\
            Chimera state & (1, $\sqrt{N}$] \\  
            Incoherence & > $\sqrt{N}$ \\
	\end{tabular}
	\label{tab:thr}
\end{table}

In the realm of complex dynamical systems, such as biophysical neural networks, the coexistence of various dimensions across spatial scales is a characteristic feature. This complexity is particularly evident in the calculation of the correlation dimension, where the dependence of $\frac{\partial \ln{C(l)}}{\partial \ln{l}}$ on $\ln l$ exhibits multiple linear segments (Fig.~\ref{fig:platos}). To address this, we propose a method for computing dimensionality within each of these segments.

The problem involves approximating the function $\frac{\partial \ln{C(l)}}{\partial \ln{l}}$ using a piece-wise constant function $\widetilde{f}$. Each interval of the constant function corresponds to a range of linear growth of $C(l)$ in log-log coordinates, allowing for the determination of characteristic dimensions of the manifold on fixed scales $l$.

Practically, one can calculate the values of $\frac{\partial \ln{C(l)}}{\partial \ln{l}}$ for a discrete set of distinct $\ln(l)$. Let $n$ represent uniformly distributed points $l_i \in [l_{\text{min}}, l_{\text{max}}]$, yielding the corresponding set of values ${y_i}: y_i = \frac{\partial \ln{C(l)}}{\partial \ln{l}}|_{l= l_i}$. This approximation problem can be addressed by minimizing the quadratic loss function $\sum_i (y_i - \widetilde{f}(l_i))^2$, a solution proposed in \citep{pmlr-v202-novikov23a}.

Figure~\ref{fig:platos} illustrates an example of distinct dimensions across spatial scales and a principal component analysis (PCA) of the corresponding dynamical regime. 

Information regarding the dimensions of the dynamical manifold on different scales enhances the understanding of system dynamics. We demonstrate its utility through an example involving a network of bursting neurons, showcasing the identification of traveling waves and various types of chimera states.

\section{Results}

To validate our approach, we investigated the dynamics of a network composed of Hindmarsh-Rose neurons \citep{hindmarsh1984model}. The system is defined by the following equations:

\begin{equation}
    \begin{cases}
        \dot x_i = ax^2_i - x_i^3 - y_i - z_i + \frac{g_{syn}}{2p}(v_R - x_i)\sum^{j=i+p}_{j=i-p}\Gamma(x_j), \\
        \dot y_i = (a + \alpha)x_i^2 - y_i, \\
        \dot z_i = c(bx_i - z_i + e), \\
    \end{cases}
    \label{eq:HR-system}
\end{equation}

where neurons are indexed by $i = 1..N$ and connected symmetrically to $p$ neighbors with a coupling strength of $g_{syn}$. The kernel of coupling is modeled by a sigmoidal function:

\begin{equation}
    \Gamma(x) = \frac{1}{1 + e^{-\lambda(x - \Theta)}}
\end{equation}

The parameter values used are: $a=2.8$, $b=9$, $c=0.001$, $e=5$, $\alpha=1.6$, $\Theta=-0.25$, $\lambda=10$, $v_R=2$.

\begin{figure}[h]
	\centering
        \includegraphics[width=1\textwidth]{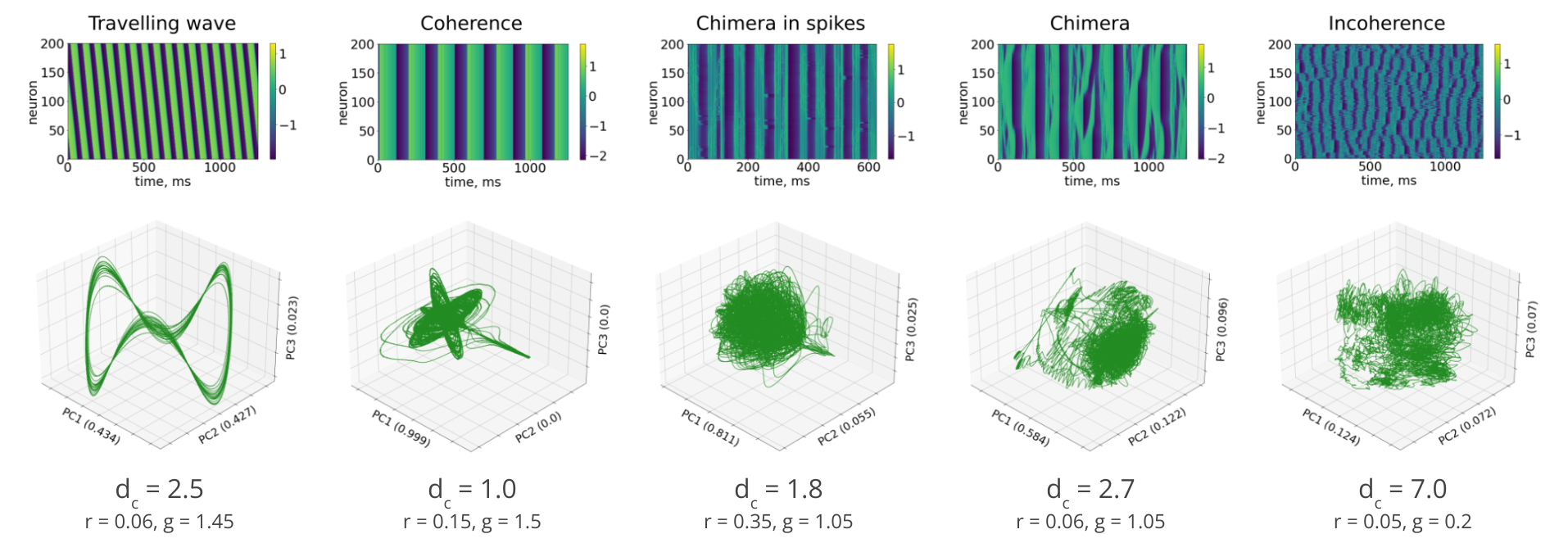} 
	\caption{Examples of dynamical regimes and correlation dimension of corresponding manifolds in the phase space. Top: spatio-temporal diagrams; Bottom: PCA-illustration of dynamical manifold of membrane potentials ($x$-variables in Eq.\ref{eq:HR-system})}
	\label{fig:examples}
\end{figure}

We initiated the analysis by identifying linear segments on the $ln C(l)$ profile. For each segment, the correlation dimension was calculated using Eq.~\ref{eq:dc}. The maximum correlation dimension was then used to distinguish between dynamical regimes (Fig.~\ref{fig:examples}). In synchronous regimes (Fig.~\ref{fig:examples}, second column), the correlation dimension ($d_c = 1$) aligns with the PCA plot, where the first principal component explains $96 \%$ of the variance. Traveling wave regimes form closed curves with dimensions closer to $2$ (Fig. \ref{fig:examples}, first column). Here, first two principal components explain $86.1\%$ of the variance. As incoherence increases, the dimensionality of the dynamical manifold grows until the system becomes fully incoherent (Fig.~\ref{fig:examples}, fifth column). For the incoherent regime, $d_c = 7$, in accordance with small values for the first PCA components (Fig.\ref{fig:examples}). 

\begin{figure}[h!]
	\centering
        \includegraphics[width=0.9\textwidth]{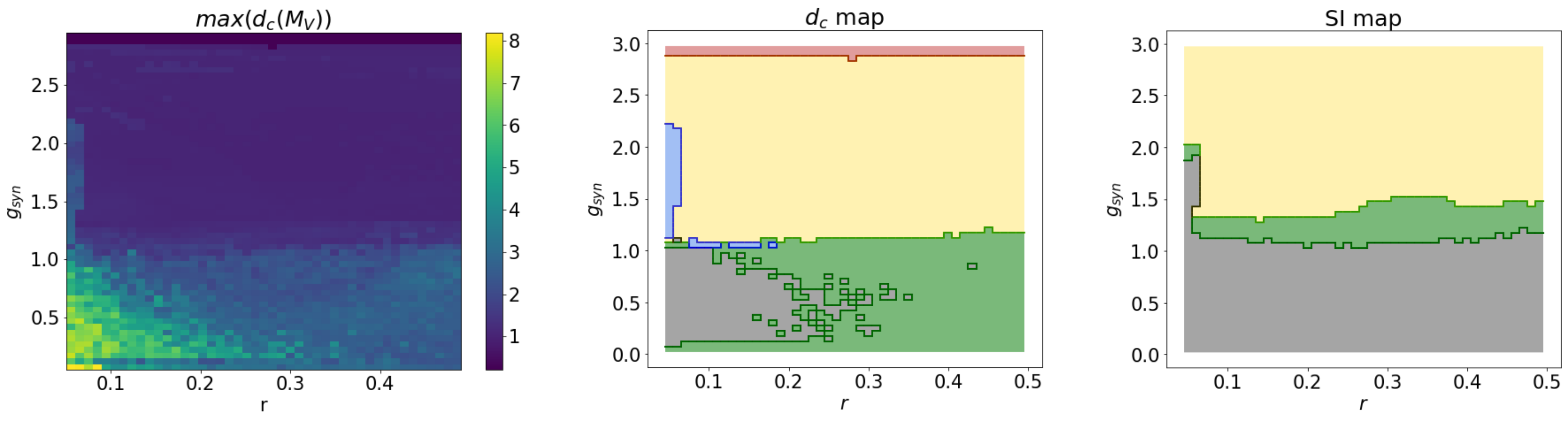} 
	\caption{Left: map of maximum $d_c$-values for dynamical manifold of membrane potential variable ($x$ in Eq.\ref{eq:HR-system}); Middle: map of dynamical regimes that is calculated using $d_c$ (Tab.\ref{tab:thr}); Right: the same map that is calculated using ground truth method, $SI$ parameter. Yellow color is for synchronous regime, green is for chimera states, blue is for travelling waves, gray is for incoherent regime and brown is for absence of oscillations. Varied parameters: $g_{syn}$ is a synaptic strength, $r$ is a connectivity. Initial conditions were randomly generated.}
	\label{fig:maps}
\end{figure}

To comprehensively explore dynamical regimes for the system defined in Eq.~\ref{eq:HR-system}, we calculated $d_c$ for each state, varying connectivity and synaptic strength (Fig.~\ref{fig:maps}A). An increase in $d_c$ is observed for small synaptic strengths and weakly connected networks. Utilizing thresholds (Tab.~\ref{tab:thr}), we constructed a map of dynamical regimes (Fig.~\ref{fig:maps}). Yellow represents synchronous regimes, green denotes chimera states, blue signifies traveling waves, gray indicates incoherent regimes, and brown marks the absence of oscillations. Parameters varied include $g_{syn}$ (synaptic strength) and $r$ (connectivity). 

We conducted a comparative analysis, between the new $d_c$ parameter against the ground truth method, Strength of Incoherence ($SI$) \citep{calim2018chimera, hizanidis2014chimera}. The $SI$ parameter is able to nearly distinguish coherence, incoherence, and chimera states (Fig.~\ref{fig:maps}, Right). However, it has limitations in detecting the absence of oscillations and traveling waves. The correlation dimension $d_c$ addresses these limitations, providing a more comprehensive understanding of the dynamical regimes under consideration.

\subsection{Two types of chimera state in bursting neurons}

\begin{figure}[h!]
	\centering
        \includegraphics[width=0.95\textwidth]{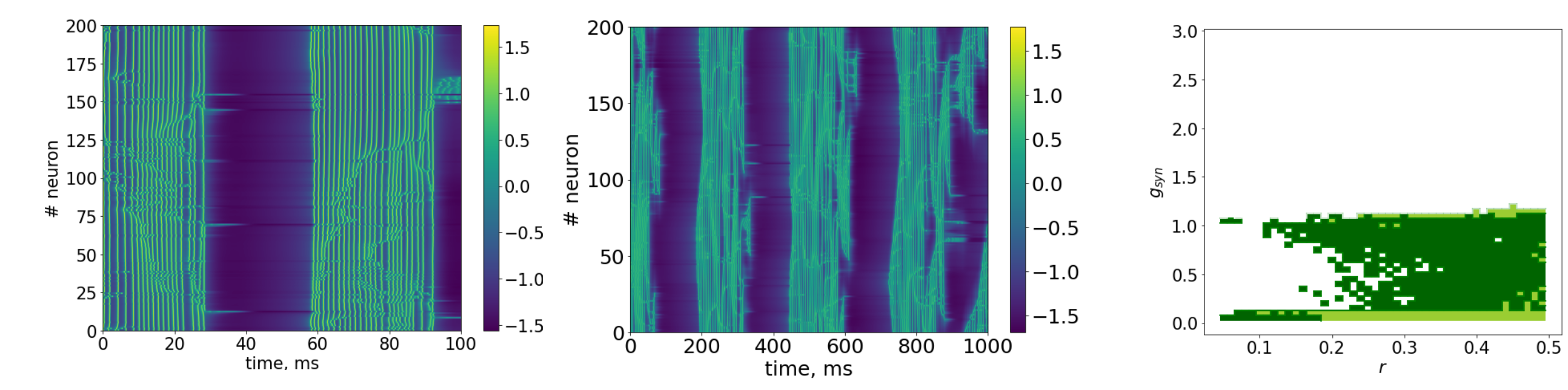} 
	\caption{Different types of chimera state: incoherence only in spikes (left) and incoherence in spikes and bursts (middle); The map of chimera states: spiking chimera is colored by light green, extensive chimera is colored by dark green (right).}
	\label{fig:diff_chimeras}
\end{figure}

The complexity of chimeras in bursting neurons lies in the separation of dynamics into two distinct scales of dynamics, slow bursts and fast spikes, and the fact that incoherence, defining partial synchronization, can also occur on these two different time-scales. In simulations, one can consider two phase subspaces corresponding to slow and fast variables. In the Hindmarsh-Rose neural network system (Eq. \ref{eq:HR-system}), each neuron is described by three variables: membrane potential ($x$), the fast spiking variable ($y$) corresponding to relatively fast ionic current, and the slow variable ($z$) corresponding to a slow adaptation current. Figure~\ref{fig:diff_chimeras} illustrates raster plots for two types of chimeras in bursting neural networks: (1) incoherence observed only between spikes within bursts, termed a spiking chimera (Fig.\ref{fig:diff_chimeras}A), (2) incoherence observed for both bursts and spikes, referred to as an extensive chimera (Fig.\ref{fig:diff_chimeras}B).

To distinguish these types of chimera states, we consider the correlation dimension of dynamical manifolds for fast ($y$ in Eq. \ref{eq:HR-system}) and slow ($z$ in Eq. \ref{eq:HR-system}) variables. In the case of a chimera in spikes, bursts are in a coherent regime while spikes have incoherent elements. This leads to the condition for a chimera in spikes: the dimension of the slow dynamical manifold $d_c^{slow} \approx 1$, while the dimension of the fast manifold $d_c^{fast} > 1$.

\begin{figure}[h]
	\centering
        \includegraphics[width=0.3\textwidth]{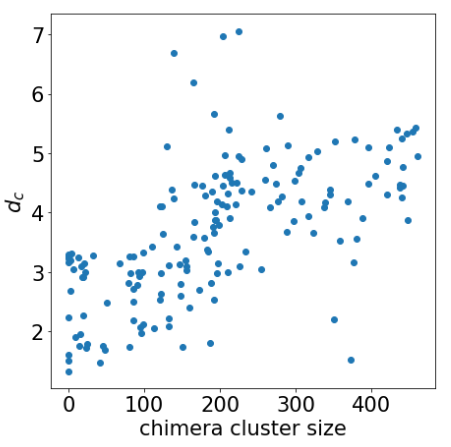} 
	\caption{Correlation dimension allows to estimate the degree of chimera which is defined as a size of incoherent cluster. The correlation is computed for spiking neural networks because there are currently no alternative reliable methods available to calculate the sizes of incoherent clusters for bursting neurons.}
	\label{fig:degree}
\end{figure}

\begin{figure}[h!]
	\centering
        \includegraphics[width=0.9\textwidth]{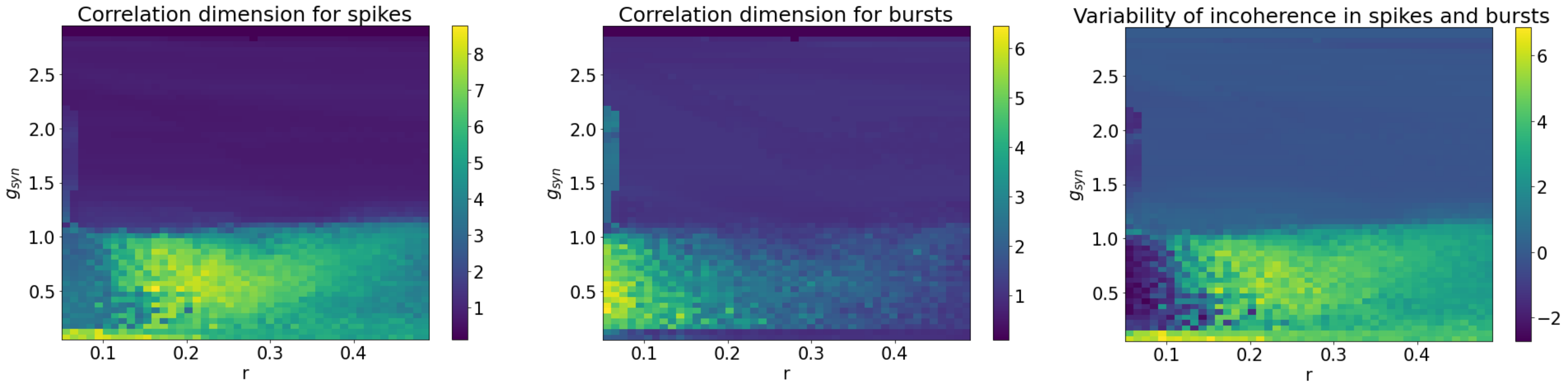} 
	\caption{Map of dynamical manifold dimensionality for fast ($y$ in Eq.\ref{eq:HR-system}) and slow ($z$ in Eq.\ref{eq:HR-system}) variables. Varied parameters: $g_{syn}$ is the synaptic strength, $r = p / N$ is the connectivity, $p$ is a number of neighbors of each neuron.}
	\label{fig:slow_fast_dc}
\end{figure}

Our assumption that there is a correlation between dimensionality and degree of chimera states (Fig.\ref{fig:degree}) due to increasing of partial synchronization manifold dimension when size of incoherent domain grows.
The degree of chimera is defined as the size of incoherent cluster. To further analyze the variability of chimera states, we utilized maps of $d_c^{fast}$ and $d_c^{slow}$, as shown in Figure \ref{fig:slow_fast_dc}. These maps provide valuable insights into the characteristics of chimera states in relation to the fast and slow components of the dynamical system.

\subsection{Distinguishing of travelling wave}

Traveling waves have become increasingly observable in brain neural activity across different spatial scales during cognitive tasks \citep{muller2018cortical, alamia2023traveling}, making this regime particularly interesting for studying in biological neural networks. To reliably identify the traveling wave regime in neural networks, we examine its distinctiveness from other dynamical states in terms of the topology of the point cloud constituting the dynamical manifold.

\begin{figure}[h]
	\centering
        \includegraphics[width=0.99\textwidth]{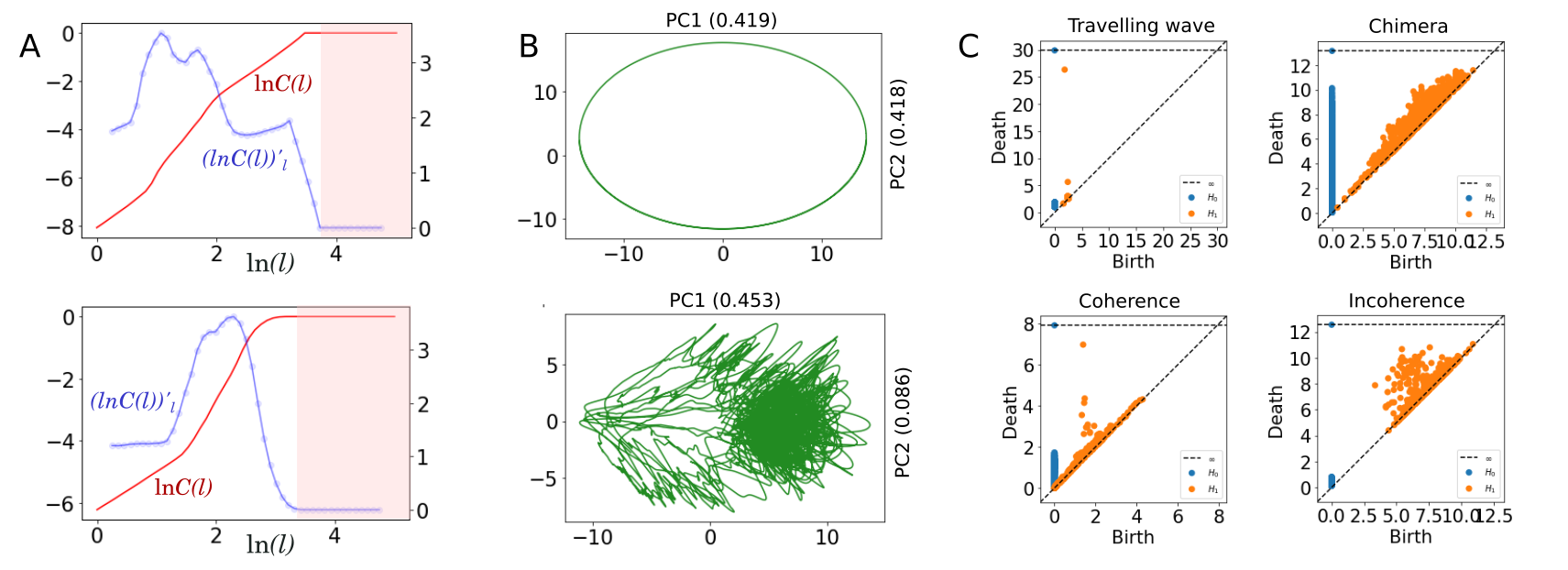} 
	\caption{An illustration of distinguishing travelling wave regime and chimera state. (A) Dependence of $\ln(C(l))$ on $l$ for travelling wave regime and chimera state; (B) 2d-PCA plots for regimes that corresponds to (A); (C) We employ topological data analysis to visualize key topological properties of travelling wave, chimera state, coherence, and incoherence. The dots represent persistent homology, with birth and death values indicating the respective "living" times as the distance between a point and the diagonal. The longer distance corresponds to a higher persistence of the corresponding homological feature throughout the filtration process.
 }
	\label{fig:tw}
\end{figure}

In a traveling wave, neurons synchronize with a gradient shift in spike phase. Two key characteristics define this regime: synchronization and phase shifts of spikes across neurons. These phase shifts generate cycles on manifolds in the phase space, as seen in the raster plot for a traveling wave in Fig.\ref{fig:examples}. Due to the synchronization, the diameter of the point cloud forming the dynamic manifold $M_V$ for the traveling wave is larger than for the chimera state. This is due to the summation of vectors along common (synchronous) directions. During synchronization, the coordinates reach their maximum simultaneously. Consequently, the diameter of the point cloud during synchronization is proportional to $N$, while during desynchronization, it is proportional to $\sqrt{N}$. This is illustrated in Fig.\ref{fig:tw}A, where $l$ with non-zero values of $\ln(C(l))$ is greater for the global synchronization regime than for the chimera state. Therefore, distinguishing the traveling wave from the chimera state involves determining the diameter of the point cloud, considering its distinction from other regimes in terms of the dynamical manifold topology.

Another topological approach worth noting that can resolve the issue of traveling wave discrimination is based on the use of persistent homology and the computation of Betti numbers \citep{gromov1981curvature}. Traveling waves exhibit stable and long-lived cycles, making them easily detectable through this method (Fig.~\ref{fig:tw}C). However, due to the computational intensity of computing Betti numbers, the analysis of the correlation dimension presents an effective and practical alternative approach in practice.

\subsection{Robustness of bursting neural network}

\begin{figure}[h!]
	\centering
        \includegraphics[width=1\textwidth]{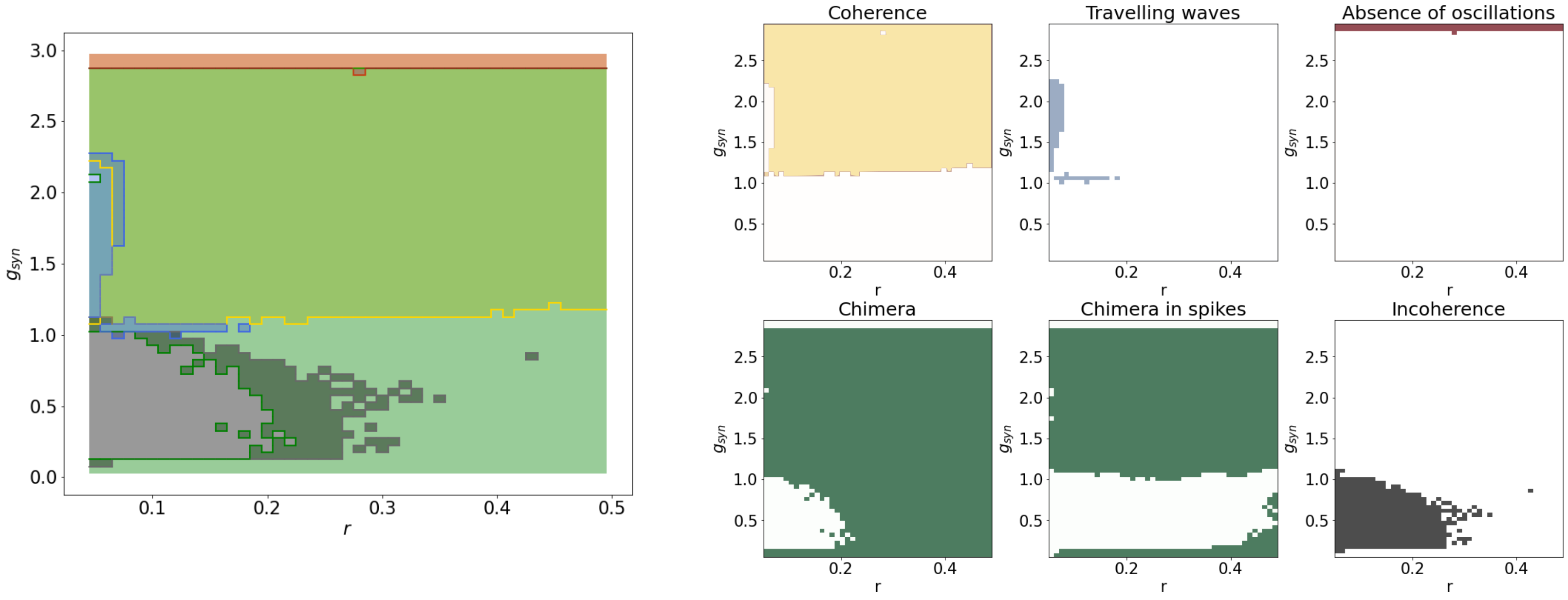} 
	\caption{The map of multistability and its separate representation as plots for different regimes in the network of HR-neurons (Eq.\ref{eq:HR-system}). Yellow color is for synchronous regime (it overlaps with a chimera state resulting into the light-green area), green is for chimera states, blue is for travelling waves, red is for the absence of oscillations and gray is for incoherent regime. Varied parameters: $g_{syn}$ is a synaptic strength, $r$ is a connectivity. }
	\label{fig:MultiMaps}
\end{figure}

The coexistence of attractors from different dynamical regimes, where each regime depends on initial conditions within different basins of attraction, is known as multistability. Utilizing the natural continuation approach proposed by \citep{dogonasheva2022multistability} and employing $d_c$ as a measure of synchronization, we identified basin boundaries and generated a map of multistability for the Hindmarsh-Rose neural network system (Eq. \ref{eq:HR-system}) (Fig.~\ref{fig:MultiMaps}).

It's noteworthy that the system of 200 burst neurons does not exhibit highly pronounced multistability. Instead, there is a predominance of regimes overlapping with chimera states, each with its stable loci. 

\section{Discussion and Conclusion}

This study addresses the intricate challenge of identifying dynamic regimes within networks of elements characterized by nontrivial dynamics. The effective control of dynamical systems necessitates thorough exploration of their parametric space, prompting the development of automated methods to discern and categorize various regimes. Current methodologies have made significant strides in distinguishing chimera states. However, their reliance on precise determinations of internal constants and limitations in discriminating between synchronization and traveling waves pose challenges.

In contrast, our proposed approach offers a more general and versatile solution, surpassing the constraints of existing methods. The dimensionality of the dynamic manifold within the phase space is independent of the signal's nature. This independence enables dimensionality methods to accurately capture and reflect underlying dynamic regimes without relying on specific system knowledge. As a result, our approach exhibits broad applicability, extending beyond spiking and bursting neurons to encompass oscillators of various natures. Its adaptability also allows for effective analysis of experimental data, given that the correlation dimension is well defined for such time series.

Our study also delved into the distinction between two types of chimera states in bursting neurons: the spiking chimera and the extensive chimera. By considering the correlation dimension of dynamical manifolds for slow and fast variables, we demonstrated the ability to discriminate between these two chimera types. This analytical approach provides a valuable tool for unraveling the complex dynamics inherent in bursting neural networks.

Moreover, our approach introduces a distinctive advantage by not only identifying chimeras but also quantifying the degree of chimera. This degree signifies the extent or magnitude of synchronization clustering observed in the system under specific parameters and initial conditions. This property enhances flexibility in controlling dynamic regimes and provides a deeper understanding of how different modulators influence the network. By presenting this novel approach, we present a powerful tool for the identification and analysis of dynamic regimes in complex networks of active elements.

Our investigation revealed a set of dynamical behaviors, illustrating the coexistence of attractors associated with different regimes. Multistability, characterized by the dependence of regimes on initial conditions within distinct basins of attraction, was identified using a natural continuation approach and the correlation dimension ($d_c$) as a measure of synchronization. The resulting map of multistability depicted regions of synchronous states, chimera states, traveling waves, the absence of oscillations, and incoherent regimes. 

It's noteworthy that the system of bursting neurons does not exhibit highly pronounced multistability. Instead, there is a predominance of regimes overlapping with chimera states, each with its stable loci. Given the prevalence of bursting neurons in the hippocampus \citep{spencer1961electrophysiology, wong1978participation}, the observed pattern suggests a more reliable coding without abrupt transitions compared to the cortex. This insight aligns with the established understanding of bursting neurons in the hippocampus \citep{zeldenrust2018neural}, emphasizing a smoother and less abrupt transition between dynamical states.

In conclusion, our study sheds light on the intricate dynamics of bursting neural networks, offering new perspectives on the identification and characterization of traveling waves, chimera states, and multistability. The correlation dimension emerges as a robust tool for distinguishing between different dynamical states, showcasing its potential for broader applications in the study of complex neural systems. Future research may delve deeper into the functional implications of the observed dynamics and explore additional topological approaches to enhance our understanding of bursting neural network behavior.

\section{Acknowledgements}
   We acknowledge Prof. N. Makarenko for valuable discussions. This publication is supported by the Brain Program of the IDEAS Research Center and Vernadski scholarship. The research was supported by RSF (project 23-22-00418) and in part through computational resources of HPC facilities at HSE University. BSG was supported by CNRS, INSERM. 

\bibliographystyle{unsrtnat}
\bibliography{main}

\end{document}